\documentclass[aps,prc,twocolumn,showpacs,floatfix,superscriptaddress,amsmath,amssymb,nofootinbib,longbibliography]{revtex4-1}

\usepackage{hyperref}
\usepackage{color}
\usepackage{amsmath}

\newcommand{\be}{\begin{equation}}
\newcommand{\ee}{  \end{equation}}
\newcommand{\ba}{\begin{eqnarray}}
\newcommand{\ea}{  \end{eqnarray}}
\newcommand{\bas}{\begin{eqnarray*}}
\newcommand{\eas}{  \end{eqnarray*}}
\newcommand{\ve}{\varepsilon}

\begin{document}

\title{Tests of the Porter-Thomas Distribution for Reduced Partial
  Neutron Widths}

\author{H.-L. \surname{Harney}}
\email{hanns-ludwig.harney@mpi-hd.mpg.de}

\author{H. A. \surname{Weidenm{\"u}ller}}
\email{haw@mpi-hd.mpg.de}

\affiliation{Max-Planck-Institut f{\"u}r Kernphysik, D-69029
  Heidelberg, Germany}

\begin{abstract}

Given $N$ data points drawn from a $\chi^2$-distribution, we use
Bayesian inference to determine most likely values and $N$-dependent
confidence intervals for the width $\sigma$ and the number $k$ of
degrees of freedom of that distribution. Using reduced partial neutron
widths measured in a number of nuclei, a guessed value of $\sigma$,
and a maximum-likelihood approach (different from Bayesian inference),
Koehler {\it et al.}~\cite{Koe10} and Koehler~\cite{Koe11} have
determined the most likely $k$-values of $\chi^2$-distributions that
fit the data. In all cases they find values for $k$ that differ
substantially from $k = 1$ (the value characterizing the Porter-Thomas
distribution (PTD) predicted by random-matrix theory). The authors
conclude that the validity of the PTD must be rejected with
considerable statistical significance. We show that the value of
$\sigma$ guessed in Refs.~\cite{Koe10, Koe11} lies far outside the
Bayesian confidence interval for $\sigma$, casting serious doubt on
the results of and the conclusions drawn in Refs.~\cite{Koe10,
  Koe11}. We also show that $\sigma$ and $k$ must both be determined
from the data. Comparison of the results with the Bayesian confidence
intervals would then decide on acceptance or rejection of the PTD.

\end{abstract}

\maketitle

\section{Introduction}

In 2010 and 2011, Koehler {\it et al.}~\cite{Koe10} and
Koehler~\cite{Koe11} have tested the Porter-Thomas distribution
predicted by random-matrix theory~\cite{Por56} against the
distribution of reduced partial neutron widths measured in a number of
nuclei. Because of possible $p$-wave admixtures and because of the
difficulty to identify narrow $s$-wave resonances unambiguously, the
authors have used a cutoff, considering only neutron resonances with
reduced partial widths larger than some energy-dependent value
$y_0$. They have applied a maximum likelihood approach to the
resonance widths so obtained and have determined the most likely value
of the parameter $k$ of a $\chi^2$-distribution with $k$ degrees of
freedom (subject to the same cutoff)~\footnote{In Refs.~\cite{Koe10,
    Koe11}, our parameter $k$ is denoted by $\nu$.}. Using $411
\ (158)$ measured partial widths in $^{194}$Pt (in $^{192}$Pt,
respectively), the most numerous sets of resonances in single nuclei,
Koehler {\it et al.}  obtain a value of $k$ close to $0.5$. They
conclude that the validity of the Porter-Thomas distribution (a
$\chi^2$-distribution with $k = 1$ degrees of freedom) must be
rejected with a statistical significance of at least $99.997$ per
cent. For the nuclear data ensemble, Koehler finds $k = 1.217 \pm
0.092$ and a statistical significance for rejection of $98.17$ per
cent~\cite{Koe11}.

Random-matrix theory forms the basis of the statistical theory of
nuclear reactions~\cite{Mit10}. Nuclear reaction cross sections that
either cannot be measured with sufficient accuracy or cannot be
measured altogether are calculated using the statistical theory. The
results are widely used in astrophysics, reactor shielding, material
science, medicine, and biology. The disagreement found in
Refs.~\cite{Koe10, Koe11} is important for fundamental and applied
science.

The results of Refs.~\cite{Koe10, Koe11} have, therefore, caused quite
a stir in the theoretical nuclear-physics community~\cite{Wei10,
  Cel11, Fyo15, Vol15, Bog17, Fan18, Fan20}. The Porter-Thomas
distribution follows from orthogonal invariance, one of the pillars of
random-matrix theory for time-reversal invariant systems. The terms
that couple an orthogonally invariant Hamiltonian to the neutron
channel and the numerous gamma channels break orthogonal invariance
and may cause deviations of the distribution of reduced partial
neutron widths from the Porter-Thomas distribution. That possibility
has been investigated in a number of papers~\cite{Cel11, Fyo15, Vol15,
  Bog17, Fan18, Fan20}. It is now firmly established, however, that
breaking of orthogonal invariance due to coupling to the channels is
by far too weak to account for the discrepancy found in
Refs.~\cite{Koe10, Koe11}, and the problem persists.

Here we address the discrepancy from a point of view that seems to
have escaped attention so far. We ask: Are the significant deviations
of the results of Refs.~\cite{Koe10, Koe11} from the Porter-Thomas
distribution caused by the limited number of partial resonance widths
available? In other words, must such deviations be expected (and not
taken as evidence against the Porter-Thomas distribution)? To answer
the question, we use the fact that each $\chi^2$-distribution is
characterized by two parameters, the number $k$ of degrees of freedom
and the width $\sigma$. We determine the minimum number of data points
(i.e., reduced partial neutron widths) needed to determine $k$ and
$\sigma$ with sufficient statistical accuracy in a sufficiently narrow
interval of values. We do so using Bayesian inference~\cite{Har16}.

The overall Bayesian approach to the $\chi^2$-distributions is
described in Section~\ref{bay1}. In Section~\ref{bay} it is applied to
estimating $\sigma$, in Section~\ref{est}, it is applied to estimating
$k$. In Section~\ref{est} we combine both results and determine the
optimum value for the pair $(\sigma, k)$. Section~\ref{sum} contains a
summary and a discussion of the implications of our work for the data
analysis of Refs.~\cite{Koe10, Koe11}. Some auxiliary calculations are
deferred to the Appendix.

\section{Bayesian Inference}
\label{bay1}

In this Section we define the $\chi^2$-distributions without and with
cutoff and the Bayesian likelihood functions for these distributions.
Given $k = 1, 2, \ldots$ independent zero-centered
Gaussian-distributed real random variables $x_l$, $l = 1, \ldots, k$
with equal variance $\sigma \neq 0$, the random variable $y$ is
defined by $y = \sum_{l = 1}^k x^2_l$. The normalized probability
distribution of $y$ is a $\chi^2$-distribution with $k$ degrees of
freedom and width $\sigma$,
\ba
\label{1}
P_{k, \sigma}(y) = \frac{1}{\Gamma(k/2) (2 \sigma^2)^{k/2}} y^{(k/2) - 1}
\exp \{ - y/(2 \sigma^2) \} \ .
\ea
Here $\Gamma(x)$ is the Gamma function~\cite{GR:15}. The normalized
$\chi^2$- distribution with cutoff at $y = y_0$ is
\ba
\label{2}
\tilde{P}_{k, \sigma}(y) &=& \Theta(y - y_0) \frac{1}{\Gamma(k/2, y_0
  / (2 \sigma)^2) (2 \sigma^2)^{k/2}} y^{(k/2) - 1} \nonumber \\
&& \qquad \times \exp \{ - y/(2 \sigma^2) \} \ .
\ea
Here $\Theta$ is the Heaviside function and $\Gamma(x, s)$ is the
upper incomplete Gamma function~\cite{GR:15}.

In general terms, our problem can be stated as follows. Given the
normalized probability distribution $P(y | \xi)$ of a variable $y$ in
terms of a parameter $\xi$, and given a number $N$ of measured values
$(y_1, y_2, \ldots, y_N)$ of $y$ (the ``data points'', jointly
referred to as ${\bf y}$), how to ascertain that $\xi$ lies with a
given probability within a given range of values? In our case, $P(y |
\xi)$ stands either for the distribution~(\ref{1}) or for the
distribution~(\ref{2}), $\xi$ stands either for $\sigma$ or for $k$,
and the data points are the measured reduced partial neutron widths.
Bayes$^\prime$ theorem~\cite{Har16} gives the answer in terms of the
``likelihood function'', i.e., the normalized posterior probability
distribution $\Pi(\xi|{\bf y})$ for $\xi$, a function of the $N$ data
points. We first display the answer for the case $N = 1$ because it is
simple and intuitively convincing. The theorem states that
\ba
\label{p1}
\Pi(\xi|y_1) = \frac{P(y_1|\xi) \mu(\xi)}{\int {\rm d} \xi \mu(\xi)
  P(y_1|\xi)} \ .
\ea
Given a data point $y_1$, the probabilty $\Pi(\xi|y_1)$ to find the
value $\xi$ equals the probability $P(y_1|\xi)$ to find, given the
parameter $\xi$, the value $y_1$ modulo two factors, the normalization
factor in the denominator of Eq.~(\ref{p1}), and $\mu(\xi)$. The
latter factor (the ``prior'') accounts for whatever {\it a priori}
knowledge we may have about $\xi$ before any data are taken. It is
possible, for instance, that $\xi$ is confined to some interval. That
knowledge, irrelevant for $P(y|\xi)$, is embodied in $\mu(\xi)$. The
generalization of Eq.~(\ref{p1}) to $N$ data points is
\ba
\label{p2}
\Pi(\xi|{\bf y}) = \frac{\mu(\xi) \ \prod_{n = 1}^N P(y_n|\xi)}{\int
  {\rm d} \xi \mu(\xi) \prod_{n = 1}^N P(y_n|\xi)} \ .
\ea
The data points are independent, and $\Pi(\xi|{\bf y})$ is the product
of their probabilities. Little is usually known about the prior
distribution $\mu(\xi)$ of $\xi$, and arguments of invariance and
symmetry are, therefore, used to determine that
quantity~\cite{Har16}. For $\xi = \sigma$, possible choices of
$\mu(\xi)$ are discussed in Section~\ref{bay}. We show that for $N \gg
1$ these choices do not influence our results.

For $N \gg 1$ the function $\Pi(\xi|{\bf y})$ tends toward a
Gaussian. That is a consequence of the central limit theorem. The
maximum yields the most likely value of $\xi$, the $N$-dependent width
of the Gaussian is used to determine the width of the interval within
which $\xi$ is found with a predetermined probability (the
``confidence interval''). The approach of $\Pi(\xi|{\bf y})$ toward
the Gaussian is not uniform in $\xi$, and corrections to the Gaussian
have to be taken into account.

In implementing that approach, we use Eq.~(\ref{p2}) to determine in
Section~\ref{bay} for fixed $k$ a range of most likely values of
$\sigma$. That range depends upon $N$. We do so for both
distributions~(\ref{1}) and (\ref{2}), without and with cutoff. In
Section~\ref{est} we proceed analogously for $k$.

\section{Bayesian Estimation of $\sigma$}
\label{bay}

For $N$ reduced neutron widths $(y_1, y_2, \ldots, y_N)$ written
jointly as ${\bf y}$, we apply, for fixed $k$, Bayes$^\prime$
theorem~(\ref{p2}) to the determination of $\sigma$. For the full
distribution~(\ref{1}), the normalized posterior distribution
$\Pi_k(\sigma|{\bf y})$ of $\sigma$ is given by
\ba
\label{4}
\Pi_k(\sigma|{\bf y}) = \frac{\mu(\sigma) \prod_{n = 1}^N
  P_{k, \sigma}(y_n)}{\int {\rm d} \sigma^\prime \ \mu(\sigma^\prime)
\prod_{n = 1}^N P_{k, \sigma^\prime}(y_n)} \ .
\ea
For the distribution~(\ref{2}) with cutoff, the normalized posterior
distribution is
\ba
\label{4a}
\tilde{\Pi}_k(\sigma|{\bf y}) = \frac{\mu(\sigma) \prod_{n = 1}^N
  \tilde{P}_{k, \sigma}(y_n)}{\int {\rm d} \sigma^\prime \ \mu(\sigma^\prime)
\prod_{n = 1}^N \tilde{P}_{k, \sigma^\prime}(y_n)} \ .
\ea
Here the cutoff may be different for each data point $y_n$. For the
prior distribution $\mu(\sigma)$ we consider two options, both
formulated as invariance requirements on the integration measure
$\mu(\sigma) {\rm d} \sigma$. (i) The measure is invariant under
translations $\sigma \to \sigma + \rho$ for all real $\rho$. That
yields $\mu(\sigma) = \mu_0$ with $\mu_0$ some real
constant\footnote{The functions~(\ref{4}), (\ref{4a}) have an
  essential singularity at $\sigma = 0$. Translational invariance can,
  nevertheless, be imposed if we replace $\sigma^2 \to \sigma^2 +
  \delta$ with positive infinitesimal $\delta$ and let $\delta \to 0$
  after the integration over $\sigma$ is done.}. (ii) The measure is
invariant under scale transformations $\sigma \to \rho \sigma$ for all
positive $\rho$. That yields $\mu(\sigma) = \mu_0 / \sigma$ with
$\mu_0$ some positive constant. We write $\mu(\sigma) = \mu_0 /
\sigma^\alpha$ with $\alpha = 0$ ($\alpha = 1$) for case (i) (case
(ii), respectively). We show that for $N \gg 1$ the two options yield
results that differ only to order $1 / N$. To leading order our
results are, thus, independent of the choice of the prior. In both
cases the integration over $\sigma^\prime$ in Eqs.~(\ref{4}, \ref{4a})
extends from $- \infty$ to $+ \infty$, and the factor $\mu_0$ cancels
in numerator and denominator of the right-hand side of these
equations.  To define $\Pi_k$ and $\tilde{\Pi}_k$ unambiguously we
replace in Eqs.~(\ref{1}) and (\ref{2}) $\sigma^k \to | \sigma^k
|$. Then $\Pi_k$ and $\tilde{\Pi}_k$ are symmetric in $\sigma$ about
the point $\sigma = 0$. The normalized distributions~(\ref{4},
\ref{4a}) are referred to as the likelihood functions for $\sigma$ for
a given data set $(y_1, y_2, \ldots, y_N)$. They are given an
additional index $\alpha = 1, 2$ to distinguish the two choices of the
invariant measure.

\subsection{Full Distribution}
\label{full}

From Eqs.~(\ref{1}), (\ref{4}) and for $\alpha = 0, 1$ we have
\ba
\label{5}
\Pi_{k, \alpha}(\sigma|{\bf y}) &=& \frac{\sigma^{- \alpha}}{{\bf
    N}_\alpha} \prod_{n = 1}^N \bigg[ \frac{1}{(2 \sigma^2)^{k/2}} \
  y^{(k/2) - 1}_n \nonumber \\
  && \qquad \times \exp \{ - y_n / (2 \sigma)^2 \} \bigg] \nonumber \\
&=& \frac{1}{{\bf N}_\alpha} \frac{1}{\sigma^\alpha} \exp \bigg\{ -
(N k/2) \ln (2 \sigma^2)
\\ && \ \ + [(k/2) - 1] \sum_{n = 1}^N \ln y_n - \sum_{n = 1}^N y_n
/ (2 \sigma^2) \bigg\} \nonumber \ .
\ea
Here ${\bf N}_\alpha$ is the normalization factor. The exponential of
$[(k/2) - 1] \sum_n \ln y_n$ does not depend on $\sigma$ and cancels
in numerator and denominator. It follows that the likelihhod
function~(\ref{5}) for $\sigma$ depends on the $N$ data points $(y_1,
y_2, \ldots, y_N)$ only via the arithmetic mean value
\ba
\label{6}
m = \frac{1}{N} \sum_{n = 1}^N y_n \ .
\ea
That is expected. We define
\ba
\label{7}
R = 2 m / k \ , \ N_0 = N k / 2 \ .
\ea
The likelihood function takes the universal form
\ba
\label{8}
\Pi_\alpha(\sigma|{\bf y}) &=& \frac{\sigma^{- \alpha} \exp \{ - N_0 [ \ln
    (2 \sigma^2) + R / (2 \sigma^2) ] \}}{\int_{- \infty}^{+ \infty}
  \frac{{\rm d} \sigma^\prime}{(\sigma^\prime)^\alpha} \exp \{ - N_0 [
    \ln (2 \sigma^{\prime 2}) + R / (2 \sigma^{\prime 2})] } \ .
\nonumber \\
\ea
The function $\Pi_\alpha(\sigma|{\bf y})$ depends on the input
parameters ($N$, $k$, $m$) only via the parameters $N_0$ and $R$
defined in Eqs.~(\ref{7}). It is convenient to introduce the
dimensionless variable
\ba
\label{9}
z = \frac{2 \sigma^2}{R} = \frac{\sigma^2}{m / k}\ , \ {\rm d}
\sigma = \frac{\sqrt{R}}{\sqrt{8 z}} {\rm d} z \ .
\ea
We absorb the factor $\sqrt{R} / \sqrt{8 z}$ into the likelihood
function. In the range $0 \leq z < \infty$ the integration measure for
$z$ is then equal to unity, and
\ba
\label{10}
\Pi_\alpha(z) &=& \frac{1}{{\cal N}_\alpha z^{(1 + \alpha)/2}} \exp \{ - N_0 [
  \ln z + 1 / z ] \} \ {\rm where} \nonumber \\
{\cal N}_\alpha &=& \int_{0}^{\infty} \frac{{\rm d} z}{z^{(1 + \alpha)/2}} \ 
  \exp \{ - N_0 [ \ln z + 1 / z ] \} \ .
\ea
As a consequence of the scaling of $\sigma$ in Eq.~(\ref{9}),
$\Pi_\alpha(z)$ depends on the data points only via the number
$N_0$. That is why ${\bf y}$ does not appear as argument of
$\Pi_\alpha(z)$. The normalization integrals ${\cal N}_\alpha$ are
calculated in the Appendix and given by
\ba
\label{11}
{\cal N}_0 &=& \frac{c(2 N_0 - 3)}{c(N_0 - 2)}
\sqrt{\frac{2 \pi}{N_0}} \exp \{ - N_0 \} \ , \nonumber \\
     {\cal N}_1 &=& \frac{c(N_0 - 1)}{\sqrt{2 \pi}} \sqrt{\frac{2 \pi}{N_0
         - 1}} \exp \{ - N_0 \} \ .
\ea
The coefficients $c(n)$ arise in the context of Stirling's formula and
for integer $n$ obey
\ba
\label{12}
\sqrt{2 \pi} \leq c(n) \leq e \ .
\ea
Combining Eqs.~(\ref{10}) and (\ref{11}) we have
\ba
\label{13}
\Pi_\alpha(z) &=& \frac{1}{{\cal C}_\alpha} \sqrt{\frac{N_0 - \alpha}
  {2 \pi}} \frac{1}{z^{(1 + \alpha)/2}} \nonumber \\
&& \qquad \times \exp \{ - N_0 [ \ln z + 1 / z - 1 ] \} \ .
\ea
The coefficients ${\cal C}_\alpha$ are given by the first factors on
the right-hand sides of Eqs.~(\ref{11}). The maximum of the term in
the exponent is at $z = 1$ or, from Eq.~(\ref{9}), at
\ba
\label{13a}
m = k \sigma^2 \ .
\ea
That shows that for $N \to \infty$, $m$ approaches asymptotically
the mean value $\langle y \rangle = k \sigma^2$ of the
distribution~(\ref{1}), as is to be expected. Writing $z = 1 + \ve$ we
expand $\Pi_\alpha(z)$ in powers of $\ve$, keeping only the
$\ve$-dependent factors and of these, only terms up to third order. We
define $\gamma_\alpha = (1 + \alpha) / (2 N_0)$. In the exponent that
gives
\ba
\label{14}
- N_0 \bigg( \gamma_\alpha \ve + \frac{1}{2} (1 - \gamma_\alpha) \ve^2
- \frac{2}{3} \ve^3 (1 - (\gamma_\alpha / 2)) \bigg) \ .
\ea
We use that in Eq.~(\ref{13}) and integrate the resulting expression
over $\ve$ from $- q$ to $+ q$. With $\zeta = \ve \sqrt{N_0 - \alpha}$
we obtain
\ba
\label{15}
&& \frac{1}{{\cal C}_\alpha \sqrt{2 \pi}} \int_{- q \sqrt{N_0 - \alpha}}
^{+ q \sqrt{N_0 - \alpha}} {\rm d} \zeta \ \exp \{ - \frac{1}{2} (1 -
\gamma_\alpha) \frac{N_0}{N_0 - \alpha} \zeta^2 \} \nonumber \\
&& \qquad \times \exp \{ - \frac{1 + \alpha}{2 \sqrt{N_0 - \alpha}}
\zeta \} \nonumber \\
&& \qquad \times \exp \{ + \frac{2}{3} \frac{N_0}{(N_0 -
  \alpha)^{3/2}}(1 - (\gamma_\alpha / 2)) \zeta^3 \} \ .
\ea
The expansion in expression~(\ref{15}) effectively proceeds in powers
of $\zeta / \sqrt{N_0}$. For $|\zeta| \leq 3$ and $N_0 = 900$ that
quantity is in magnitude smaller than or equal to $1 / 10$. In
relation to the term of second order, the third-order term in
expression~(\ref{15}) is also of order $1 / 10$. Terms of higher order
need be taken into account only for smaller values of $N_0$. The
inequalities~(\ref{12}) allow for a range of values of the
coefficients ${\cal C}_\alpha$. The least stringent bounds on $N_0$
are obtained by using the upper bounds on $1 / {\cal C}_\alpha$. These
are $1 / {\cal C}_1 \leq 1$ and $1 / {\cal C}_0 \leq e / \sqrt{2
  \pi}$. We use the first of these, putting ${\cal C}_1 = 1$. The
second actually overestimates $1 / {\cal C}_0$. By definition, the
integral~(\ref{15}) is bounded from above by unity. Numerical
calculation shows that this condition is met only for $1 / {\cal C}_0
= 1$. In effect, we replace both coefficients ${\cal C}_\alpha$ by
unity.

The integral~(\ref{15}) yields the probability $p$ to find $z$ in the
interval $(1 - q, 1 + q)$ as a function of the parameters $q$ and
$N_0$. Conversely, choosing $p$ and the said interval we may use the
integral to find $q$ and a minimum value of $N_0$. For instance, for
$\alpha = 0$, $N_0 = 900$ and $\sqrt{N_0} q = 3$, the interval is $0.9
\leq z \leq 1.1$ and $p = 0.99923$. For $N_0 = 900$ and $\sqrt{N_0} q
= 2.9$, we get $0.903 \leq z \leq 1.097$ and $p = 0.9980$. Very
similar results are obtained for $\alpha = 1$, confirming our claim
that for $N_0 \gg 1$ the Bayesian prior has little influence on our
results. It may be desirable to reduce the size of the interval so
that $z$ differs from unity (and $\sigma^2$ from $m / k$) by less than
$5$ per cent, but to keep the probability $p$ fixed. That is the case
for $\sqrt{N_0} q = 2.9$ and $N_0 = 3600$. The interval for $z$ can
also be reduced by reducing $q$. Doing so and keeping $N_0$ fixed
reduces $p$. One quickly reaches a value for $p < 1$ for which the
probability $(1 - p)$ for $z$ to lie outside the interval $(1 - q, 1 +
q)$ is unacceptably large.

We use these figures for $k = 1$, $N = 2 N_0$, the case of central
interest in Refs.~\cite{Koe10, Koe11}. The maximum of the distribution
is at $\sigma^2 = m$. For $N = 1800$, $\sigma^2$ is found with
probability $0.99923$ in the interval $0.9 \leq \sigma^2 / m \leq 1.1$
and with probability $0.9980$ in the interval $0.903 \leq \sigma^2 / m
\leq 1.097$. For $N = 400$, close to the case of $^{194}$Pt in
Ref.~\cite{Koe10}, $\sigma^2$ is found with probability $0.99923$ in
the interval $0.785 \leq \sigma^2 / m \leq 1.214$.

\subsection{Distribution with Cutoff}
\label{dis}

In view of the close similarity of the results obtained for the two
choices of the invariant measure in Section~\ref{full}, we confine
ourselves here to the case of translational invariance, put $\alpha =
0$, and drop the index $\alpha$ in what follows. We assume that the
cutoff $y_0$ is fixed, i.e., is the same for all data points $(y_1,
y_2, \ldots, y_N)$. In principle, it is possible to allow for
different cutoff values $y_{0, n}$, one for each data point. We have
not found a way, however, to handle that case analytically and
completely. At the end of Section~\ref{disc} we demonstrate, however,
that the results obtained for fixed cutoff are actually representative
of the more general case with different cutoff values.

We define the mean value $M$ of the data points as in Eq.~(\ref{6})
but display explicitly the fact that by definition all data points
obey $y_n > y_0$. Thus,
\ba
\label{6a}
M = \frac{1}{N} \sum_{n = 1}^N y_n \Theta(y_n - y_0) \ .
\ea
The normalized Bayesian likelihood function~(\ref{4a}) is given by
\ba
\label{e1}
\tilde{\Pi}_k(\sigma| {\bf y}, y_0) &=& \exp \bigg\{ - (N k / 2)
\ln (2 \sigma^2) - (N M / (2 \sigma^2)) \nonumber \\
&& - N \ln \Gamma(k/2, y_0 / (2 \sigma^2)) \bigg\} \ .
\ea
The upper incomplete Gamma function in Eq.~(\ref{e1}), defined in
Sect.~8.350 no.~2 of Ref.~\cite{GR:15}, depends upon $\sigma$. That
changes the form of the probability distribution~(\ref{8}) for
$\sigma$ in a non-universal way. The maximum is obtained by putting
the derivative with respect to $2 \sigma^2$ equal to zero. That
condition can be written as (see Sect.~8.358 no.~4 of
Ref.~\cite{GR:15})
\ba
\label{e2}
M = k \sigma^2 + (2 \sigma^2) \frac{[y_0 / (2 \sigma^2)]^{k/2}
  \exp \{ - y_0 / (2 \sigma^2) \}}{\Gamma(k/2, y_0 / (2 \sigma^2))} \ .
\ea
The right-hand side of Eq.~(\ref{e2}) is equal to the expectation
value $\langle y \rangle$ of $y$ with respect to the
distribution~(\ref{2}). Hence $M = \langle y \rangle$, as expected.
However, the last term on the right-hand side of Eq.~(\ref{e2}) is
positive. Hence, $M > k \sigma^2$. Actually, the cutoff changes
neither $\sigma^2$ nor the validity of Eq.~(\ref{13a}). However, that
equation is no longer useful for determining the most likely value of
$\sigma^2$ because only data points $y_n > y_0$ are available, and the
cutoff-dependent implicit Eq.~(\ref{e2}) must be used instead. In
general, Eq.~(\ref{e2}) cannot be solved analytically. Likewise, the
behavior of the Bayesian likelihood function near the maximum cannot
be displayed analytically for all values of $k$. We confine ourselves
here to $k = 2$ and $k = 4$ where that is possible. As shown in the
Appendix, for even values of $k$ the upper incomplete Gamma function
is given by
\ba
\label{16}
\Gamma(k/2, y_0 / (2 \sigma^2)) = \exp \{ - y_0 / (2 \sigma^2) \}
      {\cal P}_{(k/2) - 1} \ ,
\ea
and Eq.~(\ref{e1}) becomes
\ba
\label{17}
&& \tilde{\Pi}_k(\sigma|{\bf y}, {\bf y_0}) = \exp \bigg\{ - (N k/2)
\ln (2 \sigma^2) \nonumber \\
&& \qquad - N (M - y_0) / (2 \sigma^2)
- N \ln {\cal P}_{(k/2) - 1}(y_0) \bigg\} \ .
\ea
Here ${\cal P}_{(k/2) - 1}$ is a polynomial of order $(k/2) - 1$ in
$y_0 / (2 \sigma^2)$, with ${\cal P}_0 = 1$ for $k = 2$ and ${\cal
  P}_1 = 1 + y_0 / (2 \sigma^2)$ for $k = 4$.

Eq.~(\ref{17}) differs from Eq.~(\ref{8}) in two respects. First, $m$
is replaced by $(M - y_0)$, i.e., each of the $y_n$'s is replaced by
$y_n - y_0$. Second, the non-universal term $N \ln {\cal P}$ affects
the location of the maximum (see Eq.~(\ref{e2})) and, in particular,
the expansion around that maximum. For $k = 2$ we have ${\cal P}_0 =
1$. Except for the replacement $m \to (M - y_0)$, the
distribution~(\ref{17}) takes the universal form~(\ref{10}) for
$\alpha = 0$. The location of the maximum of the distribution is
shifted from $\sigma^2 = m / 2$ to $\sigma^2 = (M - y_0) / 2$. All
other conclusions of Section~\ref{full} remain unaltered. For $k = 4$
we have ${\cal P}_1 = 1 + y_0 / (2 \sigma^2)$, and the exponent in
Eq.~(\ref{17}) becomes
\ba
\label{18}
- N \bigg\{ \ln (2 \sigma^2) + \ln (2 \sigma^2 + y_0) + \frac{
  (M - y_0)}{2 \sigma^2} \bigg\} \ .
\ea
In analogy to Eq.~(\ref{9}) and since $k = 4$ we define
\ba
\label{19}
2 \sigma^2 = \frac{1}{2} (M - y_0) z \ , \ \beta = \frac{2 y_0}{m
  - y_0} \ .
\ea
In expression~(\ref{18}) we omit terms independent of $z$ generated by
the variable transformation~(\ref{19}). Expression~(\ref{18}) becomes
\ba
\label{20}
-N \bigg\{ \ln z + \ln (z + \beta) + \frac{2}{z} \bigg\} \ .
\ea
Putting the derivative equal to zero yields a quadratic equation. The
solution that tends to unity for $\beta \to 0$ is
\ba
\label{21}
z_0 = \frac{1}{2} (1 - \frac{1}{2} \beta) + \frac{1}{2} \sqrt{(1 -
  \frac{1}{2} \beta)^2 + 4 \beta}
\ea
and yields $4 \sigma_0^2 = (M - y_0) z_0$ at the maximum of
expression~(\ref{20}). We write $z = z_0(1 + \ve)$ and expand
expression~(\ref{20}) in powers of $\ve$, omitting the constant term
and keeping terms up to third order. We obtain
\ba
\label{22}
&& - N \frac{1}{2} \ve^2 \bigg( \frac{4}{z_0} - 1 - \frac{z_0^2}{(z_0
  + \beta)^2} \bigg) \nonumber \\
&& \qquad + \frac{N}{3} \ve^3 \bigg( \frac{6}{z_0} - 1
- \frac{z_0^3}{(z_0 + \beta)^3} \bigg) \ .
\ea
For $\beta = 0, z_0 = 1$ the big round brackets are equal to $2$ and
$4$, respectively, and expression~(\ref{22}) is equal to
expression~(\ref{14}) for $k = 4$, $\gamma_0 = 0$. We estimate the
range of $\beta$ as function of the cutoff using Eq.~(\ref{19}). By
construction, $M > y_0$. For the data of Ref.~\cite{Koe10} (see Figure
3) we have $M \approx 2 y_0$. Thus, a reasonable estimate for the
range of $\beta$ is $0 \leq \beta \leq 2$. In that range $z_0$ ranges
from $1$ to $\sqrt{2}$, the first big round bracket in
expression~(\ref{22}) ranges from $2$ to $\approx 1.66$, and the
second big round bracket from $4$ to $\approx 3.17$. At the upper end
of the ranges these values are equivalent to the replacement $N \to
0.8 N$. That increases the lower bound on $N$ by a factor $1.2$. Thus,
for $k = 2$ the results of Section~\ref{full} on the required minimum
size of the data set are not affected by the cutoff while for $k = 4$
the cutoff marginally increases the lower bound on the required size
of the data set.

For odd $k$ the analogue of Eq.~(\ref{16}) involves error functions
which can only be handled numerically. However, the
$\chi^2$-distribution~(\ref{1}) depends on $k$ analytically and
smoothly. Hence, there is no reason to doubt that the results obtained
for $k = 2$ and $k = 4$ hold similarly also for $k = 1$ and $k = 3$.
We conclude that the bounds on $p$ and $N$ derived in
Section~\ref{full} apply similarly also for the distribution with
cutoff. The main difference is the value of $\sigma$ for large $N$.
We have shown in Section~\ref{full} that without cutoff we have
$\sigma^2 \to m / k$ for large $N$. In the present Section, we have
shown that with a uniform cutoff at $y_0$ and for $k = 2$, we have
$\sigma^2 \to (M - y_0) / 2$, a significant change compared to
$\sigma^2 \to m / 2$ without cutoff. Although the case $k = 1$ with
cutoff is not accessible analytically, the case $k = 2$ makes us
expect that here, too, the limiting value of $\sigma^2$ differs
markedly from $m$.

\subsection{Discussion}
\label{disc}

The application of Bayes' theorem to the determination of the
parameter $\sigma$ in the $\chi^2$-distribution~(\ref{1}) for fixed
$k$ is, in principle, simple and straightforward. The likelihood
function for $\sigma$ possesses a single maximum. As the number $N$ of
data points becomes large, the location of the maximum becomes
independent of $N$ and of the Bayesian prior. It then defines the most
likely value of $\sigma^2$ in terms of the expectation value $\langle
y \rangle$. For $N \gg 1$, the likelihood function assumes the form of
a Gaussian. The width of the Gaussian is proportional to $1 /
\sqrt{N}$ and yields an estimate of the probability to find $\sigma^2$
within some narrow interval centered at the maximum. For realistic
values of $N$ (i.e., for a few hundred data points), the Gaussian
approximation is replaced by the analytical form of the normalized
likelihood function. Integration of that function over an interval of
$\sigma$-values centered at the maximum yields the probability $p$ to
find $\sigma$ within that interval. For fixed $p$ the size of the
interval shrinks with increasing $N$. Fixing $p$ and the size of the
interval determines the minimum number $N$ of data points.

In Section~\ref{full} we have implemented that program for the
$\chi^2$-distribution~(\ref{1}) without cutoff. The most likely value
of $\sigma$ is given by $\sigma^2 = m / k$. That result is expected
because for $y = \sum_{l = 1}^k x^2_l$ defined as in
Section~\ref{bay1} in terms of a sum of real Gaussian-distributed
zero-centered random variables $x_l$ with equal variance $\sigma^2$,
we obviously have $\langle y \rangle = k \sigma^2$. The Bayesian
approach defines a confidence interval for $\sigma^2$ in terms of $N$,
or vice versa. For useful values of $p$ and acceptable values of the
size of the interval, we find $N_0 \geq 900$. Eq.~(\ref{7}) yields $N
= 2 N_0 / k$. For fixed $N_0$, the bound on $N$ decreases with
increasing $k$. That is because the $\chi^2$-distribution~(\ref{1}) is
most asymmetric for $k = 1$. For the Porter-Thomas distribution ($k =
1$), we obtain $N \geq 1800$.

The implementation of that program for the
$\chi^2$-distribution~(\ref{2}) with cutoff in Section~\ref{dis}
encounters an additional problem. The relation $\sigma^2 = m / k$ is
useless because $m$ as defined in Eq.~(\ref{6}) cannot be determined
from the data. The mean value $M$ defined in Eq.~(\ref{6a}) in terms
of the data points that are actually available is surely bigger than
$m$ so that $M = c \ m$ with $c \geq 1$. The relation $\sigma^2 = m /
k$ remains intact but is now written as $\sigma^2 = M / (c k)$. Here
$c$ depends upon $k$ and the cutoff $y_0$. Qualitatively, $c$ depends
on these parameters as follows. For $k \gg 1$ the maximum of the
$\chi^2$-distribution~(\ref{1}) shifts to ever larger $k$-values, and
the cutoff becomes increasingly irrelevant. Therefore we have $c \to
1$ for $k \to \infty$. The converse is true for $k \to 0^+$. The bulk
of the $\chi^2$-distribution becomes ever more narrowly concentrated
near zero. In fact, the relation $\sigma^2 = m / k$ shows that $m \to
0$ for $k \to 0^+$. On the other hand, $M$ attains by definition a
finite value $\geq y_0$ in that limit. Hence $c \to \infty$ for $k \to
0$. We conclude that $c \geq 1$ diverges for $k \to 0$ and decreases
monotonically with increasing $k$. For fixed $k$, $c$ is expected to
grow monotonically with $y_0$. The most likely value for $\sigma^2$ of
the Bayesian likelihood function in Eq.~(\ref{e2}) takes these facts
into account.  Because of the occurrence of the incomplete Gamma
function, that equation cannot be solved analytically and in general
for $\sigma^2$ .  We have confined ourselves to determining the most
likely value of $\sigma^2$ for $k = 2$ and $k = 4$ and for a fixed
value of $y_0$ as only these cases can be worked out analytically. For
$k = 2$, the most likely value of $\sigma^2$ is $\sigma^2 = (M - y_0)
/ 2$. For $y_0 = (1/2) M$ that value differs by the factor $(1/2)$
from the value $\sigma^2 = m / k$ without cutoff. For $k = 4$, the
most likely value is $\sigma^2 = z_0 (M - y_0) / 4$. For $y_0 = M / 2$
we have $z_0 = \sqrt{2}$ (see Eq.~(\ref{21})) and obtain $\sigma^2 = M
\sqrt{2} / 8$. That value differs by the factor $\sqrt{2} / 2 \approx
0.7$ from the value $\sigma^2 = m / 4$ without cutoff. For $k = 1$ we
expect an even smaller value $\sigma^2 \approx 0.4 M$. For the bounds
on $N$, the modifications due to the cutoff are less drastic and
amount to an increase of $20$ per cent or so.

When a separate cutoff $y_{0, n}$ is used for each data point $y_n$,
Eq.~(\ref{e2}) is replaced by
\ba
\label{e3}
M &=& k \sigma^2 \\
&& + (2 \sigma^2) \frac{1}{N} \sum_{n = 1}^N \frac{[y_{0, n} /
    (2 \sigma^2)]^{k/2} \exp \{ - y_{0, n} / (2 \sigma^2) \}}
{\Gamma(k/2, y_{0, n} / (2 \sigma^2))} \ . \nonumber
\ea
Comparison of Eqs.~(\ref{e3}) and (\ref{e2}) suggests defining $y_0$
by putting
\ba
\label{e4}
&& (2 \sigma^2) \frac{[y_0 / (2 \sigma^2)]^{k/2} \exp \{ - y_0 /
  (2 \sigma^2) \}}{\Gamma(k/2, y_0 / (2 \sigma^2))} \nonumber \\
&& = (2 \sigma^2) \frac{1}{N} \sum_{n = 1}^N \frac{[y_{0, n} /
    (2 \sigma^2)]^{k/2} \exp \{ - y_{0, n} / (2 \sigma^2) \}}
{\Gamma(k/2, y_{0, n} / (2 \sigma^2))} \ .
\ea
The second line in that equation can be read as an average over the
cutoff parameters $y_{0, n}$. It comprises only positive
contributions. Moreover, each term in the sum increases monotonically
with increasing $y_{0, n}$. Therefore, Eq.~(\ref{e4}) possesses a
unique solution $y_0$. That solution determines the factor $c$
reducing $M / k$. Thus, the conclusions drawn above on the dependence
of $c$ on the cutoff parameter $y_0$ remain valid. The confidence
interval for $\sigma$ cannot be determined by the solution $y_0$ of
Eq.~(\ref{e4}) but must be determined from all the $y_{0, n}$. It is
reasonable to expect, however, that the estimates in Section~\ref{dis}
for that interval remain valid.

\section{Bayesian Estimation of $k$}
\label{est}

In Section~\ref{bay} we have defined the minimum number of data points
needed to determine for fixed $k$ and required accuracy the parameter
$\sigma$ in the $\chi^2$-distributions $P_{k, \sigma}(y)$ of
Eq.~(\ref{1}) and $\tilde{P}_{k, \sigma}(y)$ of Eq.~(\ref{2}). With
the help of Bayes' theorem, we now proceed analogously for the
parameter $k$. For $N$ data points $(y_1, y_2, \ldots, y_N)$ (written
below as ${\bf y}$) and for a fixed value of $\sigma$ we determine the
most likely value of $k$ and a confidence interval for that parameter.
At the end of the Section we combine both approaches to determine the
optimum value for the pair $(k, \sigma)$ of parameters that
characterize the $\chi^2$-distributions in Eqs.~(\ref{1}) and
(\ref{2}) and the accuracy with which these can be determined. As in
Section~\ref{bay} we address first the full
$\chi^2$-distribution~(\ref{1}) and then the distribution~(\ref{2})
with cutoff.

For the $\chi^2$-distribution of Eq.~(\ref{1}), the normalized
Bayesian likelihood function $\Pi(k|{\bf y})$ is
\ba
\label{23}
\Pi(k|{\bf y}) = \frac{\mu(k) \ \prod_{n = 1}^N P_{k, \sigma}(y_n)}
   {\int {\rm d} k \ \mu(k) \ \prod_{n = 1}^N P_{k, \sigma}(y_n)} \ .
\ea
We consider $k$ as a continuous variable in the range $0 < k <
\infty$. A meaningful invariance requirement on the prior distribution
$\mu(k)$ is scale invariance so that $\mu(k) = \mu_0 / k$. Then the
integral in the denominator extends from $0$ to $\infty$, and $\mu_0$
cancels in numerator and denominator. We use Eq.~(\ref{1}) and write
Eq.~(\ref{23}) as
\ba
\label{24}
&& \frac{k^{- 1}}{\cal N} [(2 \sigma^2)^{1/2} \Gamma(k/2)]^{- N}
\exp \{ - N m / (2 \sigma^2) \} \nonumber \\
&& \qquad \times \exp \bigg\{ [(k / 2) - 1] \sum_n \ln [y_n / (2
  \sigma^2)]  \bigg\} \ .
\ea
Here ${\cal N}$ is the normalization factor. That factor is well
defined because the Gamma function diverges sufficiently rapidly for
both $k \to \infty$ and $k \to 0^+$. The factors $(2 \sigma^2)^{-
  N/2}$ and $\exp \{ - N m / (2 \sigma^2) \}$ are independent of $k$,
cancel in numerator and denominator, and are omitted in what
follows. The likelihood function $\Pi(k|{\bf y})$ depends on the $N$
data points only via the parameter
\ba
\label{25}
{\cal L} = \frac{1}{N} \sum_n \ln [y_n / (2 \sigma^2)] \ .
\ea
For fixed $\sigma^2$, ${\cal L}$ is determined by the geometric mean
value of the data points. The $N$-dependent terms in the numerator of
$\Pi(k|{\bf y})$ are
\ba
\label{26}
\exp \{ - N \ln \Gamma(k/2) + N [(k/2) - 1] {\cal L} \} \ .
\ea
The function of $k$ in the exponent has a maximum at $k = k_0$ where
\ba
\label{27}
{\cal L} &=& \frac{\rm d}{{\rm d} x} \ln \Gamma(x)\bigg|_{x = (k_0/2)}
\ .
\ea
For $x \to 0^+$ (for $x \to + \infty$), the logarithmic derivative of
the Gamma function, written as $f'(x)$, approaches $- \infty$ ($+
\infty$, respectively). In the interval $0^+ \leq x < \infty$, $f'(x)$
increases monotonically with $x$. Therefore, Eq.~(\ref{27}) always has
a uniquely defined solution $k_0$. The accuracy with which $k_0$ is
determined depends on the probability to find $k = k_0 ( 1 + \ve)$
within some interval centered at $k_0$. That interval is obtained by
expanding the exponent in expression~(\ref{26}) in powers of $\ve$. We
write $f''(k/2), f'''(k/2)$ for the higher-order logarithmic
derivatives of the Gamma function. Keeping only terms of second and
third order in $\ve$ we find
\ba
\label{30}
- \frac{N k_0}{2} \bigg( \bigg[ \frac{1}{2} k_0 f''(k_0/2) \bigg]
\frac{\ve^2}{2} + \bigg[ \frac{1}{4} f'''(k_0/2) k^2_0 \bigg]
\frac{\ve^3}{6} \bigg) \ .
\ea
To calculate expression~(\ref{30}) we use Sect.~8.362, no.~1 of
Ref.~\cite{GR:15},
\ba
\label{28}
f'(x) = - \frac{1}{x} - \gamma - \sum_{n = 1}^\infty \bigg( \frac{1}
{n + x} - \frac{1}{n} \bigg) \ .
\ea
Here $\gamma \approx 0.577216$ is Euler's constant. That gives
\ba
\label{31}
f''(x) &=& \frac{1}{x^2} + \sum_{n = 1}^\infty \frac{1}{(n + x)^2} \ ,
\nonumber \\
f'''(x) &=& - \frac{2}{x^3} - \sum_{n = 1}^\infty \frac{2}{(n + x)^3} \ .
\ea
We evaluate expression~(\ref{30}) for two representative values $k_0 =
1$ and $k_0 = 2$ of $k_0$. With $x = (k_0/2)$ Eqs.~(\ref{31}) yield
\ba
\label{32}
f''(1/2) &=& 4 \sum_{n = 0}^\infty \frac{1}{(2 n + 1)^2} \approx
4.9300 \ , \nonumber \\
f'''(1/2) &=& - 16 \sum_{n = 0}^\infty \frac{1}{(2 n + 1)^3} \approx
- 16.8288 \ , \nonumber \\
f''(1) &=& \sum_{n = 0}^\infty \frac{1}{(n + 1)^2} = \frac{\pi^2}{6}
= 1.6400 \ , \nonumber \\
f'''(1) &=& - 2 \sum_{n = 0}^\infty \frac{1}{(n + 1)^3} \approx
- 2.4042 \ .
\ea
Expression~(\ref{30}) becomes
\ba
\label{33}
{\rm For} \ k = 1 &:& \ - \frac{N}{2} ( 1.23 \ve^2 - 0.70 \ve^3) \ ,
\nonumber \\
{\rm For} \ k = 2 &:& \ - \frac{N}{2} ( 1.64 \ve^2 - 0.80 \ve^3) \ .
\ea
These figures are quite close to each other indicating that the
constraints on $N$ derived from Eqs.~(\ref{33}) hold similarly for all
values of $k$ in the interval $1 \leq k \leq 2$ and beyond. In
Section~\ref{bay} we have used the full analytical knowledge of the
likelihood function in combination with an expansion as in
expression~(\ref{30}) to determine the interval confining
$\sigma^2$. That cannot be done here because the normalization
integral involves the Gamma function and is not available
analytically. We resort to the rule~\cite{Har16} usually applied in
that case. We request that the term of third order in
expression~(\ref{30}) be less than or equal in magnitude to $1 / 10$
times the term of second order everywhere in the interval $(-3,
+3)$. When used for the figures in Section~\ref{full}, that rule
yields $N \geq 1070$, close to the value of $N_0 = 900$ obtained
there. For $k = 1$ and expression~(\ref{33}) that gives $N \geq 180$
and $| \ve | \leq 0.17$. For $N = 400$ (representative for
Ref.~\cite{Koe10}) we find $| \ve | \leq 0.11$, for $N = 50$
(representative for Ref.~\cite{Koe11}) we find $| \ve | \leq 0.31$.
In all cases, the upper bound on $\ve$ defines the interval $(1 - \ve)
\leq 1 \leq (1 + \ve)$ confining $k$. The expansion used in
expression~(\ref{30}) is determined entirely by the logarithmic
derivative of the Gamma function and independent of the actual value
of $\sigma^2$ and so are, therefore, the bounds on $k$ just obtained.

With cutoff and for fixed $\sigma$, the Bayesian likehood function for
the distribution~(\ref{2}) as function of $k$ is
\ba
\label{34}
\tilde{\Pi}(k|{\bf y}) = \frac{\mu(k) \ \prod_{n = 1}^N
  \tilde{P}_{k, \sigma}(y_n)}{\int {\rm d} k \ \mu(k) \ \prod_{n = 1}^N
  P_{k, \sigma}(y_n)} \ .
\ea
We again assume that the cutoff $y_0$ is the same for all data points
$(y_1, y_2, \ldots, y_N)$ all of which obey $y_n > y_0$. Then the
numerator in Eq.~(\ref{34}) is
\ba
\label{35}
&& \mu(k) [(2 \sigma^2)^{1/2} \Gamma(k/2, y_0 / (2
  \sigma^2))]^{- N} \exp \{ - N m / (2 \sigma^2) \} \nonumber \\
&& \qquad \times \exp \{ N ((k/2) - 1){\cal L} \} \ .
\ea
The incomplete upper Gamma function disappears rapidly for $k \to
\infty$ but takes a finite value for $k \to 0^+$. Therefore, the
normalization integral in Eq.~(\ref{34}) exists only if we choose
$\mu(k) = \mu_0$. We do so even though the problem lacks translational
invariance. The analogue of Eq.~(\ref{27}) is
\ba
\label{36}
{\cal L} &=& \frac{\rm d}{{\rm d} x} \ln \Gamma(x, y_0 / (2 \sigma^2))
\bigg|_{x = (k_0/2)} \ .
\ea
The logarithmic derivative of the incomplete upper Gamma function has
a large negative value at $k = 0$ and diverges for $k \to
\infty$. Therefore, a solution $k_0$ of Eq.~(\ref{36}) always exists.
The accuracy with which $k_0$ is determined is defined in the same
manner as in Eqs.~(\ref{30}) to (\ref{33}). Details depend upon the
actual choice of $y_0$ and are not investigated here. We only remark
that to account for the cutoff, it is probably convenient to write
$\Gamma(k/2, y_0 / (2 \sigma^2)) = \Gamma(k/2) - \gamma(k/2, y_0 / (2
\sigma^2))$ and to use Eqs.~(\ref{31}) for $\Gamma(x)$, see
Sect.~8.350, no.~1, of Ref.~\cite{GR:15}. For the incomplete lower
Gamma function $\gamma(x, s)$ powerful series expansions are
available. In any case there is no reason to assume that the results
will differ significantly from those of the case without cutoff.

Combining the result of Section~\ref{bay} with the present one is
straightforward without cutoff. In Section~\ref{full} the most likely
value for $\sigma$ was found to be $\sigma = m / k$. That yields $N
{\cal L} = \sum_n \ln (k y_n / 2 m)$, and Eq.~(\ref{27}) becomes
\ba
\label{29}
\frac{1}{N} \sum_n \ln (y_n / m) + \ln (k_0 / 2) = f'(k_0/2) \ .
\ea
Eq.~(\ref{29}) allows for a direct determination of $k_0$ from the
input parameters. The uncertainty of $\sigma^2$ values due to the
finite number of data points causes via Eq.~(\ref{25}) the same
uncertainty (percentagewise) of ${\cal L}$. We estimate the resulting
uncertainty in $k_0$ using Eq.~(\ref{27}). The derivative of the
right-hand side of that equation equals $f''(k_0/2)$. At $k = 1$, that
has the value $4.93$. Thus, the uncertainty of $k_0$ is about $1 / 5$
of that of $\sigma^2$. That must be combined with the uncertainty in
$k_0$ estimated below expressions~(\ref{33}). For the data in
Ref.~\cite{Koe10}, the uncertainty in $\sigma^2$ (in $k_0$) amounts to
$20$ percent ($10$ percent, respectively), giving a total uncertainty
for $k_0$ of about $11$ percent.

With cutoff, the situation is more complex. For small $k$ we lack an
analytical expression for the most likely value of $\sigma^2$. For $k
\gg 1$ we may use $\sigma^2 = m / k$ as before. But for $k \to 0$,
$\langle y \rangle$ tends toward a nonzero value (and not to zero, as
the relation $k \sigma^2 = m$ would suggest). For purposes of
orientation we may put $M = (k + a) \sigma^2$ and fit $a$ to the
result for $k = 2$ in Section~\ref{dis}. That gives $M = k \sigma^2 +
2 y_0$. Here $M$ is given by Eq.~(\ref{6a}). We may use that
expression for $\sigma$ in Eq.~(\ref{36}). That yields a first guess
for $k_0$. For higher accuracy, it is necessary to solve
Eq.~(\ref{e2}) numerically. We have shown, in any case, that for $k =
1$, $\sigma^2$ is equal to or smaller than about $0.4 \ M$. The
influence of the cutoff on the uncertainty of $\sigma^2$ and $k_0$ is
probably less strong. For $\sigma^2$ that was shown in
Section~\ref{bay}. For $k_0$ it follows because in the interval $0
\leq k \leq \infty$, the derivatives of the Gamma function and of the
upper incomplete Gamma function are, for cutoff values $y_0 / (2
\sigma^2) \approx 1 / 2$, quite close to each other except at the
lower end points.

\section{Summary and Conclusions}
\label{sum}

The $\chi^2$-distributions in Eqs.~(\ref{1}) and (\ref{2}) depend on
the number $k$ of degrees of freedom and on the parameter $\sigma$
characterizing the width of the distributions. We have used
Bayes$^\prime$ theorem to determine these parameters from a set of $N$
data points $(y_1, y_2, \ldots, y_N)$. The maxima of the two Bayesian
likelihood functions, probability distributions for $k$ and $\sigma$,
respectively, define the most probable values of these parameters.
They yield $k$ in terms of the logarithm of the geometric mean value
of the data points and as a function of $\sigma$, and $\sigma$ as the
arithmetic mean value $m$ of the data points and as a function of
$k$. The combination of both yields unique values for $k$ and
$\sigma$. The width of each of the two Bayesian likelihood functions
around their maxima depends on the number $N$ of data
points. Requesting that $k$ lies with a given probability within a
given interval centered at the maximum of its Bayesian likelihood
function, fixes the minimum number $N$ of data points needed for $k$,
and correspondingly for $\sigma$. Conversely, the number $N$ of data
points determines the confidence intervals for $k$ and $\sigma$. As a
rule of thumb we have shown that for realistic values of $N$, the
confidence interval for $\sigma^2$ ($k$) has a width of about $\pm 20$
percent ($\pm 10$ percent, respectively) around the maximum.

The results for $k$ obtained in Refs.~\cite{Koe10, Koe11} lie outside
the range for $k$ determined above for the Porter-Thomas distribution,
roughly given by $0.89 \leq k \leq 1.11$. That seems to confirm the
rejection of the Porter-Thomas distribution in Refs.~\cite{Koe10,
  Koe11}. However, in Refs.~\cite{Koe10, Koe11} the parameter
$\sigma^2$ in the $\chi^2$-distribution~(\ref{2}) with cutoff is
replaced by the value $\sigma^2 = M / k$, and the result is used in a
maximum-likelihood analysis with the aim to find the value of $k$ that
best fits the data. Here $M$ is the arithmetic average of the data
points all of which are bigger than the cutoff. That replacement is
incorrect because $M$ is bigger by a factor $c > 1$ than $m$, the
arithmetic mean without cutoff. The correct relation is $\sigma^2 = m
/ k = M / (c k)$. That reduces by $1 / c$ the value of $\sigma^2$ used
in Refs.~\cite{Koe10, Koe11}. For $k = 4$ and $k = 2$ the factor $c$
was found to be about $1.3$ and $2.0$, respectively. For $k = 1$ we
expect $c \approx 2.5$. That is a major change which is likely to
affect the analysis of Refs.~\cite{Koe10, Koe11} in a substantial
manner and casts doubt on the results and the conclusions.

Why did Koehler {\it et al.} and Koehler choose a value for $\sigma^2$
that is so far outside the Bayesian confidence interval?
Refs.~\cite{Koe10} and \cite{Koe11} use the $\chi^2$-distribution for
$k$ degrees of freedom written as in Eq.~(1) of Ref.~\cite{Koe11}. In
our notation and for $s$-wave resonances that equation reads
  \ba
  \label{koe}
  P_{k} &=& \frac{k}{2 \Gamma(k/2)} \bigg(\frac{k y}{2 \langle y
    \rangle}\bigg)^{(k/2) - 1} \exp \bigg\{ - \frac{k y}{2 \langle y
    \rangle} \bigg\} \ .
  \ea
Eq.~(\ref{koe}) is obtained from Eq.~(\ref{1}) upon replacing
$\sigma^2 \to \langle y \rangle / k$. While Eq.~(\ref{1}) holds
generally, such replacement and the resulting Eq.~(\ref{koe}) hold
only under two restrictive conditions. (i) The number $N$ of data
points is large. The resulting Bayesian confidence interval for
$\sigma^2$ is so narrow that $\sigma^2$ may be replaced by the most
likely value. (ii) The most likely value of $\sigma^2$ is given by
$\langle y \rangle / k$. Condition (i) is not met by the data sets in
Refs.~\cite{Koe10} and \cite{Koe11}. That is addressed in the
paragraph that follows.  More importantly, condition~(ii) holds for
the case without cutoff (see Eq.~(\ref{13a})) but only for that
case. With cutoff, $\sigma^2$ is determined by the implicit
Eq.~(\ref{e2}). We have shown that the result differs markedly from
$\langle y \rangle / k$. Using Eq.~(\ref{koe}) as the starting point
of a maximum-likelihood analysis as done in Refs.~\cite{Koe10} and
\cite{Koe11} for the case with cutoff is flawed from the
beginning. Combined with the replacement $\langle y \rangle \to M$ it
implies the guess $\sigma^2 = M / k$ which is incorrect.
  
Replacing $\sigma^2$ in the $\chi^2$-distribution by some fixed value
entails a second problem. Such replacement is correct only in the
limit $N \to \infty$ where the maximum of the Bayesian likelihood
function is completely sharp. For finite $N$ the Bayesian likelihood
function defines an $N$-dependent interval of acceptable values for
$\sigma^2$. That interval may be fairly wide. For $^{194}$Pt we have
$N = 411$\cite{Koe10}. For $N = 400$ and $k = 1$ Eq.~(\ref{7}) yields
$N_0 = 200$. Putting $q \sqrt{N_0} = 2.9$ as at the end of
Section~\ref{full}, we obtain a probability $p = 0.9980$ for
$\sigma^2$ to differ by less than $\pm 20$ percent from its most
likely value. For the data in Ref.~\cite{Koe11} the size of the
$\sigma^2$-interval increases by factors that lie between $4.5$ and
$2$. A set of $N$ data points corresponds to a pair $(\sigma^2, k)$ of
values that lie somewhere in these intervals. Different sets of $N$
data points determine different pairs $(\sigma^2, k)$ of values. It
is, therefore, incorrect to determine $k$ by choosing arbitrarily some
value of $\sigma^2$ within the given interval. Both $\sigma^2$ and $k$
must be determined from the given data set. That can be done, for
instance, with the help of the maximum-likelihood analysis of
Refs.~\cite{Koe10, Koe11}. Within the intervals defined by $N$, that
determines to which $\chi^2$-distribution the data belong.

Using the method of Ref.~\cite{Koe10} for a simultaneous determination
of $\sigma^2$ and of $k$ may be realistic for the Pt
isotopes~\cite{Koe10} where the number of data points is large. The
Nuclear Data Ensemble~\cite{Koe11} comprises $24$ nuclides. In all but
two of these, the number $N$ of resonances is between $17$ and
$109$. For each nuclide, $\sigma^2$ and $k$ must be determined
independently. That is necessary because the mean value of the input
parameters $y_n$ (and, therefore, $\sigma^2$) depends upon the neutron
strength function which varies with mass number. Thus, application of
the method of Ref.~\cite{Koe10} to the Nuclear Data Ensemble poses a
challenge.

In summary, the analysis of Refs.~\cite{Koe10, Koe11} is
unsatisfactory as it stands. The results are inconclusive. A
reevaluation is needed that determines both $\sigma^2$ and $k$ from
the data and compares the result with confidence intervals derived
from the Bayesian analysis. Only on that basis will it be possible to
decide whether the data support the Porter-Thomas distribution, or
whether that distribution must be rejected.

{\bf{Acknowledgement}}. We acknowledge useful correspondence by
P. Koehler.

\section*{Appendix: Normalization Integrals}

The normalization integral ${\cal N}_\alpha$ in Eq.~(\ref{9}) is
calculated separately for $\alpha = 0$ and $\alpha = 1$. For $\alpha =
0$ we substitute $z = N_0 x^{- 2}$, $(1/z) {\rm d} z = - (2 \sqrt{N_0}
/ x^2) {\rm d} x$. That yields
\ba
\label{A1}
&& {\cal N}_0 = \int_0^\infty \frac{{\rm d} z}
{\sqrt{z}} z^{- N_0} \exp \{ - N_0 / z \} \nonumber \\
&& \qquad = N_0^{(1/2) - N_0} \int_{- \infty}^{+\infty} {\rm d} x \
(x^2)^{N_0 - 1} \exp \{ - x^2 \} \nonumber \\
&& \qquad = N_0^{(1/2) - N_0} (-)^{N_0 - 1} \nonumber \\
&& \qquad \qquad \times \frac{\partial^{N_0 - 1}}{\partial
  \beta^{N_0 - 1}} \int_{- \infty}^{+\infty} {\rm d} x
\ \exp \{ - \beta x^2 \} \bigg|_{\beta = 1} \nonumber \\
&& \qquad = \sqrt{\pi} N_0^{(1/2) - N_0} (-)^{N_0 - 1}
\frac{\partial^{N_0 - 1}}{\partial \beta^{N_0 - 1}}
\beta^{- (1/2)} \bigg|_{\beta = 1} \nonumber \\
&& \qquad = 2 \sqrt{\pi N_0} (2 N_0)^{- N_0} (2 N_0 - 3)!! \ .
\ea
We use Stirling's formula,
\ba
\label{A2}
&& n! = c(n) \ n^{n + (1/2)} \exp \{ - n \} \ , \ {\rm with} \nonumber \\
&& \sqrt{2 \pi} \leq c(n) \leq e \ {\rm for \ all} \ n \ .
\ea
That gives
\ba
\label{A3}
&& (2 N_0 - 3)!! = \frac{(2 N_0 - 3)!}{2^{N_0 - 2} (N_0 - 2)!} \\
&=& \frac{c(2 N_0 - 3)}{c(N_0 - 2)} \frac{(2 N_0 - 3)^{2 N_0 - 3 + (1/2)}}
      {2^{N_0 - 2} (N_0 - 2)^{N_0 - (3/2)}} \exp \{ 1 - N_0 \} \nonumber \\
      &\approx& \frac{c(2 N_0 - 3)}{c(N_0 - 2)} \exp \{ 1
      - N_0 \} \nonumber \\
      && \times \exp \{ (N_0 - 1) \ln N_0 + (N_0 - (1/2)) \ln 2 - 1 \}
      \ . \nonumber
\ea
The normalization integral becomes
\ba
\label{A4}
{\cal N}_0 &=& \frac{c(2 N_0 - 3)}{c(N_0 - 2)}
\sqrt{\frac{2 \pi}{N_0}} \exp \{ - N_0 \} \ .
\ea

For $\alpha = 1$ we substitute $z = N_0 / x$, ${\rm d} z = - (N_0 /
  x^2) {\rm d} x$ and obtain, using Stirling's formula~(\ref{A2}),
\ba
\label{A5}
      {\cal N}_1 &=& N_0^{- N_0} \int_0^\infty {\rm d} x \ x^{N_0 - 1}
      \exp \{ - x \} \nonumber \\
      &=& N_0^{- N_0} (N_0 - 1)! \approx \frac{c(N_0 - 1)}{\sqrt{N_0 - 1}}
      \exp \{ - N_0 \} \ .
\ea

For $k$ even and $s = (k/2) - 1 = 0, 1, 2, \ldots$ the upper
incomplete Gamma function is given by
\ba
\label{A6}
&& \Gamma(k/2, y_0 / (2 \sigma^2)) = \int_{y_0 / (2 \sigma^2)}^\infty
t^s \exp \{ - t \} \ {\rm d} t \nonumber \\
&& \qquad = (-)^s \frac{\partial^s}{\partial \beta^s} \frac{1}{\beta}
\exp \{ - \beta y_0 / (2 \sigma^2) \} \bigg|_{\beta = 1} \nonumber \\
&& \qquad = {\cal P}_s \exp \{ - y_0 / (2 \sigma^2)
\} \ .
\ea
Here ${\cal P}_s$ is a polynomial in $y_0 / (2 \sigma^2)$ of order
$s$, with
\ba
\label{A8}
{\cal P}_0 = 1 \ , \ {\cal P}_1 = 1 + y_0 / (2 \sigma^2) \ .
\ea

\end{document}